\documentclass[conference]{IEEEtran}
\IEEEoverridecommandlockouts
\usepackage{cite}
\usepackage{amsmath,amssymb,amsfonts}
\usepackage{algorithmic}
\usepackage{graphicx}
\usepackage{textcomp}
\usepackage{xcolor}
\def\BibTeX{{\rm B\kern-.05em{\sc i\kern-.025em b}\kern-.08em
    T\kern-.1667em\lower.7ex\hbox{E}\kern-.125emX}}

\usepackage[normalem]{ulem}

\usepackage{url}

\usepackage{booktabs}
 \usepackage{array,multirow,graphicx}

\usepackage{color,soul}

\begin{document}

IEEE Copyright Notice
Copyright (c) 2022 IEEE
Personal use of this material is permitted.  Permission from IEEE must be obtained for all other uses, in any current or future media, including reprinting/republishing this material for advertising or promotional purposes, creating new collective works, for resale or redistribution to servers or lists, or reuse of any copyrighted component of this work in other works.\\

\noindent This is the \emph{author-submitted article} (see the IEEE Preprint Policy \url{https://cis.ieee.org/publications/t-emerging-topics-in-ci/tetci-ieee-preprint-policy} (accessed 4 July 2022)). The \emph{accepted article} will be published in:\\

\noindent Proceedings of the 2022 IEEE International Conference on Decentralized Applications and Infrastructures (DAPPS’22: 15-18 Aug 2022, San Francisco, USA)\\
DOI: to be published

\newpage

\title{Can We Effectively Use Smart Contracts to Stipulate Time Constraints?}

\newcommand{\code}[1]{\mbox{\texttt{#1}}}
\newcommand{\R}{\mathbb{R}}
\newcommand{\N}{\mathbb{N}}

\author{
	
	\IEEEauthorblockN{Tobias Eichinger}
	\IEEEauthorblockA{\textit{Service-centric Networking} \\
		\textit{Technische Universität Berlin}\\
		Berlin, Germany \\
		tobias.eichinger@tu-berlin.de\\
		0000-0002-8351-2823}
	
	\and
	
	\IEEEauthorblockN{Marcel Ebermann}
	\IEEEauthorblockA{\textit{Service-centric Networking} \\
		\textit{Technische Universität Berlin}\\
		Berlin, Germany \\
		marcel.ebermann@tu-berlin.de}
}

\maketitle

\begin{abstract}
Smart contracts provide the means to stipulate rules of interaction between mutually distrustful organizations. They encode contractual agreements on the basis of source code, which else need to be contractualized in natural language. While the mediation of contractual agreements via smart contracts is seamless in theory, it requires that the conditions of an interaction are accurately made available in the blockchain. Time is a prominent such condition. In the paper at hand, we empirically measure the consistency of a smart contract to yield equal results on the basis of the time of an interaction and its potentially inaccurate representation in the blockchain. We propose a novel metric called \emph{execution accuracy} to measure this consistency. We specifically measure the execution accuracy of a time interval-constrained smart contract that executes distinct logic within and without some constraint interval. We run experiments for the local \emph{Ganache} and \emph{Quorum} and the public \emph{G\"{o}rli} and \emph{Rinkeby} Ethereum blockchains. Our experiments confirm our intuition that execution accuracy decreases near interval bounds. The novelty of our proposed metric resides in its capacity to quantify this decrease and make distinct blockchains comparable with respect to their capacity to accurately stipulate time contraints.\\
\end{abstract}


\maketitle

\section{Introduction}
\label{sec:introduction}

Smart contracts provide the means to stipulate rules of interaction between mutually distrustful organizations. They have been proposed to execute business \cite{Mendling2018bcbpm,Weber2016monitoring} and manufacturing \cite{Nielsen2020digitaltwin,Westerkamp2020tracing} processes seamlessly across multiple organizations. Although conceptually seamless, a practical difficulty is the accurate \emph{time injection} into a blockchain, where we denote by time injection the act of making world time available for smart contract execution \cite{Muehlberger2020}. Consider for instance a smart contract that periodically updates a variable holding the current world time as a UNIX timestamp. As part of the blockchain state, world time can now be used within smart contracts. 

Time is a key aspect of business and manufacturing processes \cite{Abid2019,Cheikhrouhou2015}. Consider for instance two manufacturing activities that require a certain time delay. Only if the start and end time of an activity are accuractely injected into a blockchain, delays and time constraints can be correctly coordinated. In particular, coordinating an entire process requires time to be injected not only once but continuously throughout the process. We denote by continuously injecting time into a blockchain as \emph{time tracing}. In brief, process execution depends on  time tracing, which in turn depends on time injection.

\begin{figure}[t]
		\centering
		\includegraphics[width=\linewidth]{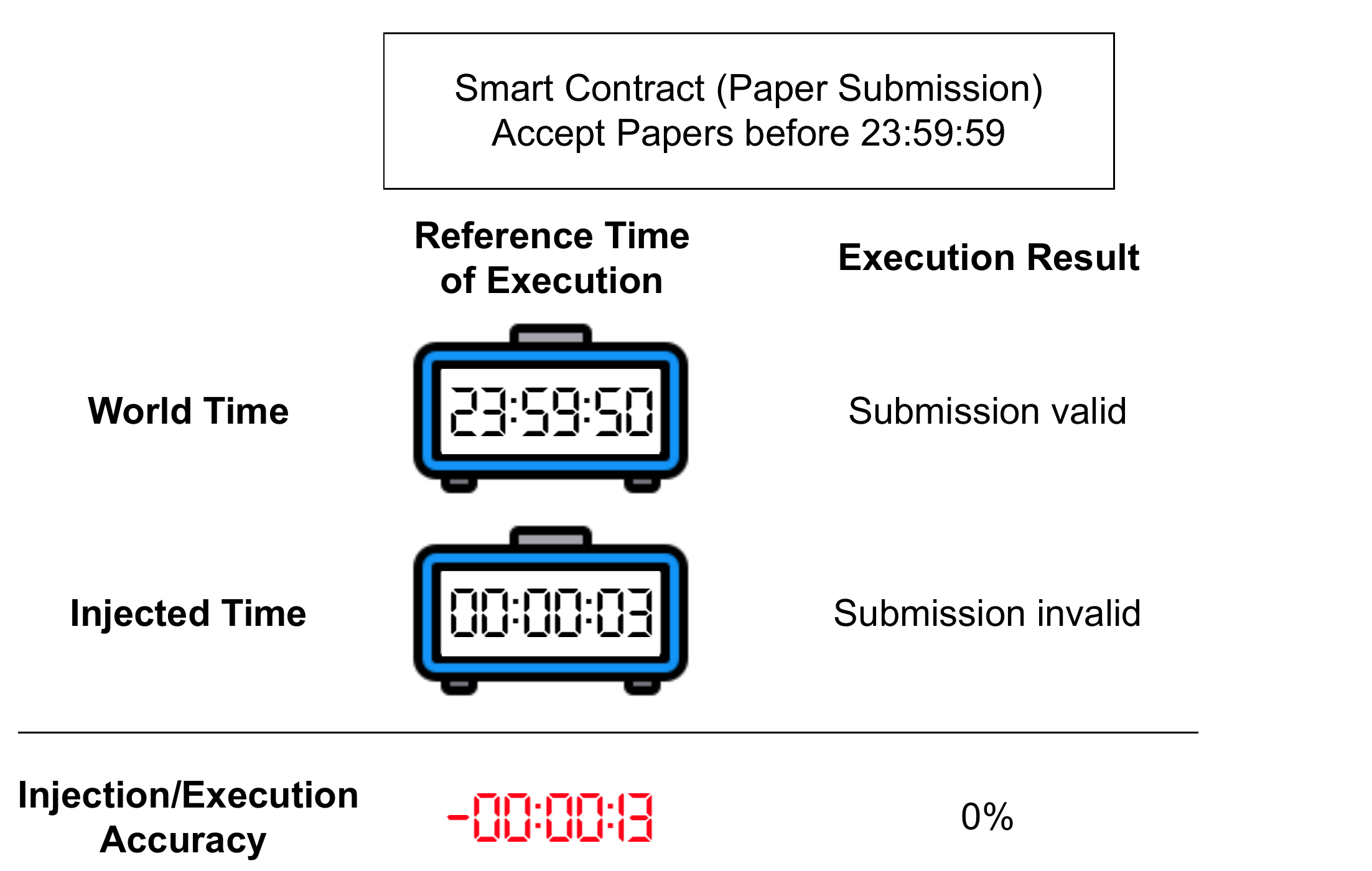}
		\caption{Comparison of injection and execution accuracy. While injection accuracy measures time offsets, execution accuracy measures probabilities.}
		\label{fig:executionaccuracy}
\end{figure}

Ladleif and Weske \cite{Ladleif2020} survey five distinct time injection methods for use in blockchains. They conclude that there is no objectively best time injection method. The so-called parameter method is ideal for scenarios in which senders of transactions are trustworthy. Here, senders of a transaction truthfully attach the current world time to a transaction. World time and injected time are hence trivially identical. In this case, the result of executing smart contract logic on the basis of world time and injected time is identical.\footnote{Here, we assume that execution logic is independent from other transactions calling the smart contract.} In contrast, if world time and injected time are not equivalent, execution may yield inaccurate results due to inaccurate time injection. Figure \ref{fig:executionaccuracy} illustrates inaccuracies caused by inaccurate time injection. 

We denote by \emph{execution accuracy} the probability that the execution of smart contract logic on the basis of world time and injected time yield identical results. The parameter method mentioned in the previous paragraph yields perfect execution accuracy. However, it is only useful in scenarios in which senders of transactions are trustworthy. If senders are not trustworthy, the \emph{block timestamp method} is an alternative time injection method \cite{Ladleif2020}. Here, world time is injected by the miner, instead of the sender. World time is injected as the block timestamp of the block the calling transaction is mined into. 

We denote by \emph{injection accuracy} the offset between world time and injected  time. The injection accuracy of the block timestamp method then depends on for instance (a) the  block time, that is the difference in timestamps of consecutive blocks, (b) the fee offered to the miners, and (c) the miners themselves who set block timestamps. While injection accuracy for the block timestamp method is well-understood \cite{Ladleif2020monitoring,Yasaweerasinghelage2017}, execution accuracy is not. This is due to the fact that injection accuracy solely depends on the network properties of the blockchain, while execution accuracy additionally depends on the specific execution logic of smart contracts.

In summary, coordinating process execution on blockchains is only meaningful if the result of calling a smart contract is consistent over world time and injected time. Injection accuracy cannot fully describe such consistency, since injection accuracy is independent from smart contract execution logic. In order to address this gap, we propose the following contributions: 

\begin{itemize}
	\item We introduce execution accuracy as a metric that measures the consistency with which time-sensitive smart contract execution on world time and injected time match. 
	\item We measure execution accuracy of a time interval-constrained smart contract with respect to the block timestamp method. Measurements are made on two local Ethereum implementations \emph{Ganache}\footnote{https://trufflesuite.com/ganache/ (accessed 5 May 2022)} and \emph{Quorum}\footnote{https://consensys.net/quorum/ (accessed 5 May 2022)} and the \emph{G\"{o}rlie}\footnote{\url{https://goerli.net/} (accessed 5 May 2022)} and \emph{Rinkeby}\footnote{\url{https://www.rinkeby.io/} (accessed 5 May 2022)} Ethereum test networks. 
	\item We share the code of our test bed with the community. 
\end{itemize} 
\section{Related Work}
\label{sec:related_work}

We first review literature that studies time injection independently from the execution logic of smart contracts with respect to their injection delay and injection accuracy. Afterwards, we tend to  application-specific aspects of time injection. 

\subsection{Accuracy and Latency of Time Injection}
\label{sec:timeinjection}

Blockchains can only validate whether time was injected correctly, that is formally 
correctly according to the blockchain protocol. Measuring injection accuracy is infeasible from within a blockchain. Injection accuracy needs to be  measured outside of it. In particular for time injection via so-called oracle contracts, measuring injection accuracy can be translated into a trust issue toward the stakeholder that injects time \cite{Ladleif2020monitoring,Muehlberger2020}. Roughly, injected time can be assumed to be accurate if the stakeholder injecting time is trustworthy. Applications typically do require continuous injection of time instead of single instances of time injection. We tend to prior work in applications that continuously inject time.

\subsection{Time Injection in Applications}
\label{sec:timetracing}

Tracing digitally on a blockchain what happens physically in the world is key to collaborative manufacturing processes \cite{Mandolla2019,Nielsen2020digitaltwin,Putz2021}. This is because smart contracts cannot immediately access the state of the physical world. In particular, they cannot immediately access world time. Consider for instance a manufacturing process that requires a product to cool down to a certain temperature before it can be prepared for delivery. The temperature is traced on the blockchain such that the manufacturing process can be correctly coordinated. Using a digital model that traces the state of a physical product is generally denoted using a \emph{digital twin} \cite{vanDerHorn2021digitaltwin}. 

While collaborative manufacturing mainly traces the state of a product, collaborative business processes typically trace interactions between stakeholders \cite{Abid2019}. The business process itself is executed in a process engine \cite{LopezPintado2019}. Here, smart contracts define conditions, particularly temporal conditions, for interactions and contractual agreements between stakeholders \cite{Abid2019,Cheikhrouhou2015}. Consider for instance a smart contract that sets the due date for the reception of products to prepare for delivery. The delivery service provider may then refuse to receive products when the smart contracts verifies that the due date has passed. 

Time tracing is distinct from time synchronization. Time synchronization aims to equalize offsets in distinct world time measurements \cite{Regnath2020}. In contrast, time tracing is about continuously persisting world time references on a blockchain. Since blockchains establish a concept of time that is distinct from world time \cite{Swan2016}, time tracing cannot be considered synchronization as no offsets are harmonized. It is important to see that block timestamps represent references to world time, they are not themselves world time timestamps. 

We now explain how block timestamps as references to world time timestamps can be injected into a blockchain and used to implement time interval constraints in smart contracts.

\section{Concept}
\label{sec:concept}

We formalize the block timestamp method. We then define a prototypical time interval-constrained smart contract. Finally, we present our test bed and describe how it can be used to measure execution accuracy.

\subsection{The Block Timestamp Method} 
\label{sec:notation}

We first introduce some notation. Let $B=(B_i)_{i \in \N}$ be an infinite sequence of blocks that represent a blockchain. Every block $B_i$ holds a sequence of transactions. Transactions in a block have been found valid by a miner, where valid means that state transitions conform with the underlying blockchain protocol. Hence, any finite subsequence of blocks $(B_i)_{1\leq i \leq N}$ represents a valid state of the blockchain $B$. We associate every block $B_i$ with a timestamp $t_i \in \R_{\geq0}$ such that $t_i < t_j$ for all $i<j$. Then blockchain $B$ is associated with an infinite and strictly increasing sequence $t = (t_i)_{i \in \N}$ of so-called block timestamps. Block timestamps are not essential to blockchains, yet convenient since they reference when a block was mined. 

We formalize the block timestamp method as described for instance in \cite{Ladleif2020}. Let $x$ be a transaction sent at world time $t_x$ that calls some time-sensitive smart contract ${\cal C}$. We denote by $\hat{t}_x$ the injected time of world time $t_x$. Let further $B_i$ be the latest block and $t_i$ its block timestamp. Then $x$ is communicated to miners within the blockchain network. Miners determine a valid next block $B_{i+1}$ that includes $x$ and is associated to some block timestamp $t_{i+1}>t_i$. Particularly, the block timestamp method sets $t_{i+1}=\hat{t}_x$. The reference time used to execute the time-sensitive smart contract ${\cal C}$ is exactly the block timestamp. We now present our test bed in which time-sensitivity represents a time interval constraint.

\begin{table}[t]
	\centering
	\caption{Four distinct states of a time interval-constrained smart contract as described in Section \ref{sec:smartcontract}.} \label{tab:confusiontable}
	\begin{tabular}{ccc}
		&$\hat{t}_x\in I$; ($\hat{p}=1$)  &$\hat{t}_x\notin I$; ($\hat{p}=0$) \\
		\midrule
		$t_x\in I$; ($p =1$)      & true positive & false negative\\
		$t_x\notin I$; ($p=0$) & false positive& true negative\\
		\bottomrule
	\end{tabular}
	\\[6pt]	
\end{table}

\subsection{Time Interval-Constrained Smart Contract}
\label{sec:smartcontract}
We define a prototypical time-sensitive smart contract that implements an interval time constraint. Similarly to the example shown in Figure \ref{fig:executionaccuracy},  the prototypical time-sensitive smart contract then executes distinct behavior within and without some reference time interval $I$. Formally, let ${\cal C}$ be a smart contract initialized with some interval  $I=[a,b]\subset \R_{\geq0}$ and two binary state variables $p$ and $\hat{p}$. Let further $x$ be some transaction submitted to the blockchain network at world time $t_x$ and included into a block $B_i$ with block timestamp $t_i$. Then ${\cal C}$ sets its state variable $p=1$ if world time $t_x$ satisfies the time constraint ($t_x \in I$), and else $p=0$ ($t_x \notin I$). Analogously, we set $\hat{p}=1$ and $\hat{p}=0$ for injected time $\hat{t}_x$. 
 
 Table \ref{tab:confusiontable} shows the four possible states defined by the binary state variables $p$ and $\hat{p}$. In analogy to binary classification in machine learning, we associate the state variable $p$ with the \emph{true condition} of world time. From the perspective of the blockchain, we interpret injected time as a prediction of the true world time when transaction $x$ was sent. We associate the state variable $\hat{p}$ with the \emph{predicted condition} of world time. Table \ref{tab:confusiontable} thus technically represents a \emph{confusion matrix}.

 Recall that execution accuracy measures the consistency with which a smart contract yields the same result with reference to world time and injected time. In other words, it measures the probability that the smart contract ${\cal C}$ enters either a true positive or true negative state. We now present the test bed we use to measure this probability.

\subsection{Test Bed}
\label{sec:experimentalsetup}

We empirically estimate the probability that a transaction $x$ submitted to the blockchain network at world time $t_x$ either yields a true positive or true negative smart contract state. Since the state of the smart contract ${\cal C}$ depends on world time $t_x$ and injected time $\hat{t}_x$, we need to make both available at execution time. We use the parameter method (see Section \ref{sec:introduction}) to make world time available and  choose another time injection method such as the block timestamp method (see Section \ref{sec:notation}) to make injected time available. The smart contract ${\cal C}$ can then determine its state after being called by a transaction. 

\begin{figure}[t]
	\centering
	\includegraphics[width=\linewidth]{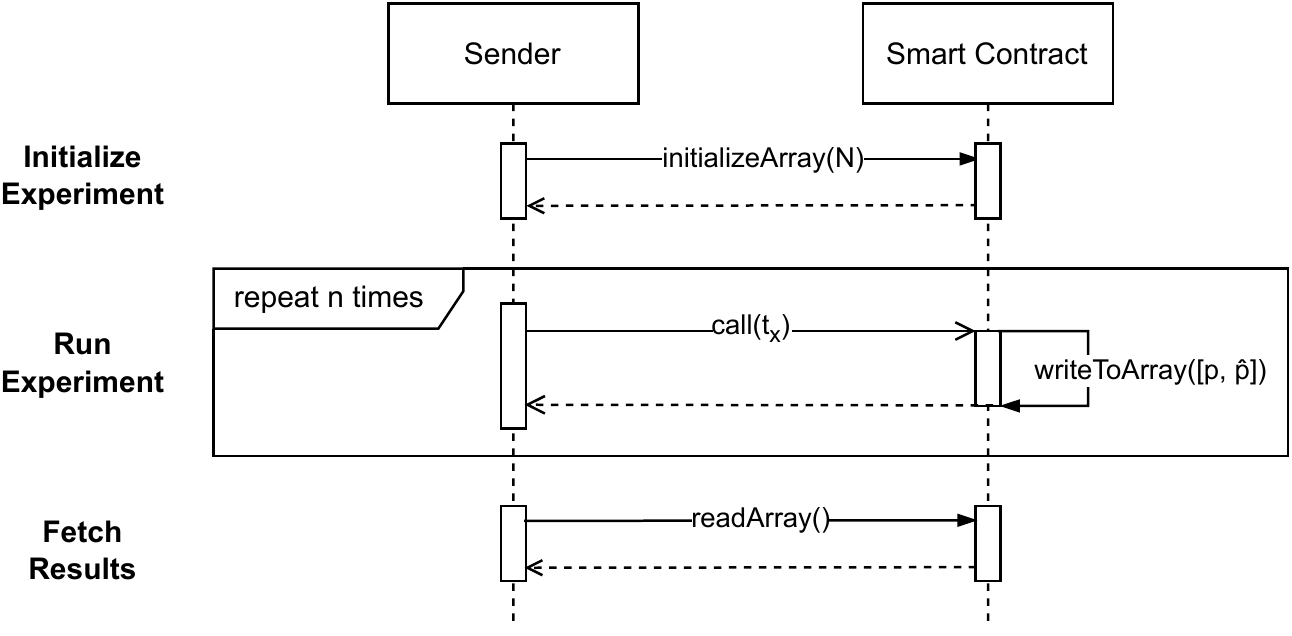}
	\caption{Experimental setup used to measure execution accuracy, where $t_x$ denotes world time when sending a calling transaction $x$ to the smart contract and $[p,\hat{p}]$ the smart contract state induced by it. For presentational simplicity, we omit the intermediary blockchain between sender and smart contract.}
	\label{fig:setup}
\end{figure}

In order to estimate probabilities for each of the four possible states to occur, we call the smart contract at regular intervals by sending a transaction to it. We send a total of $N$ transactions over an experiment interval  $J\supset I$ that includes the constraint interval  $I=[a,b]$ of the time-sensitive smart contract ${\cal C}$. Note that sending a transaction to ${\cal C}$ overwrites any previous state of ${\cal C}$. In order to persist all state changes over the course of the experiment, we first initialize an array of length $n$ in the smart contract. Throughout the experiment, we then persist all $n$ state changes ($[p,\hat{p}]$) in the array. After the experiment, we fetch the results by reading the filled array from the smart contract. Figure \ref{fig:setup} shows a schematic workflow of an experiment. It remains to formally define execution accuracy and how to read it off the array.    

\begin{figure*}[t]
	\centering
	\includegraphics[width=\linewidth]{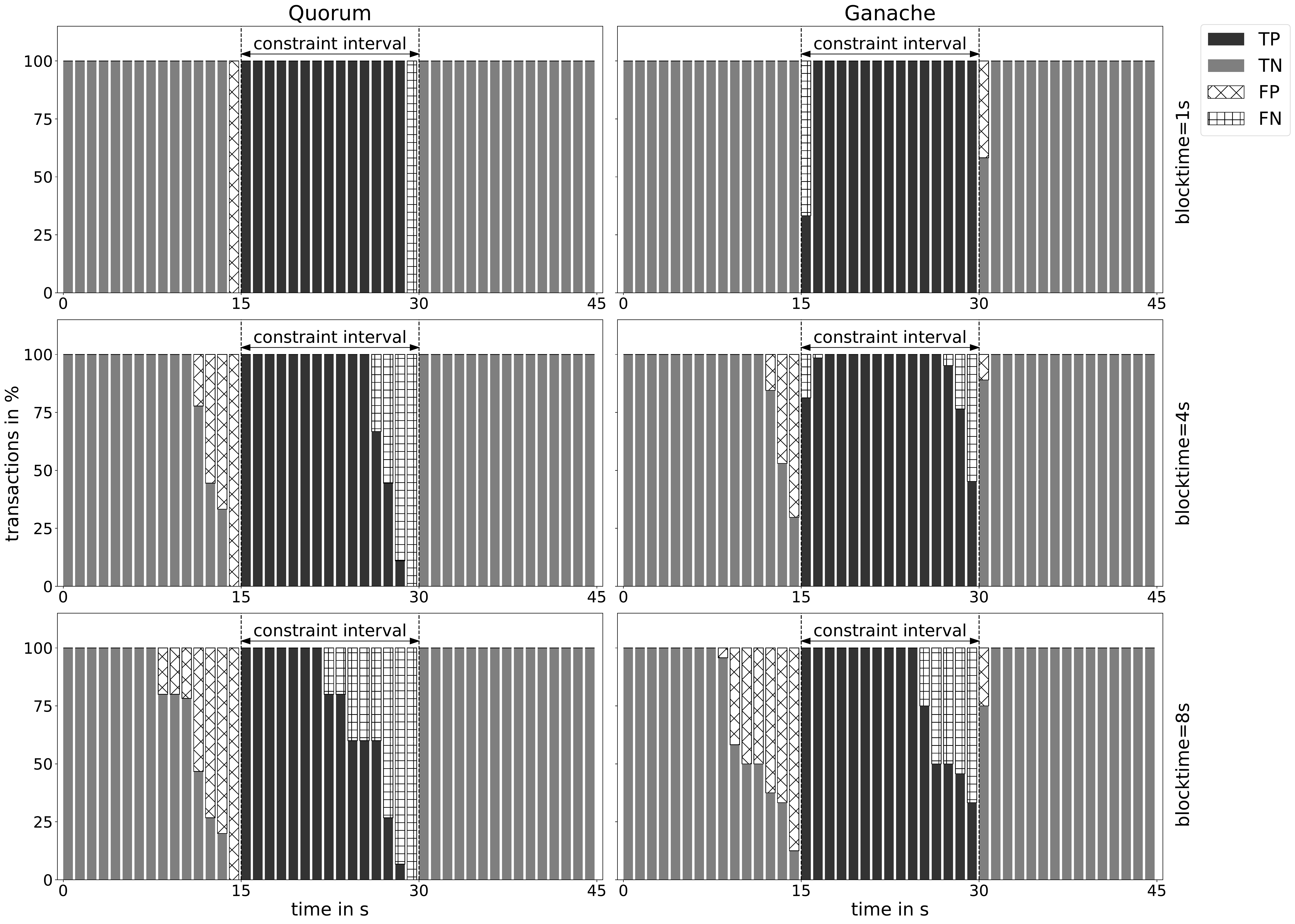}
	\caption{Relative frequencies of true positives (TP), true negatives (TN), false positives (FP), and false negatives (FN) for a time interval-constrained smart contract and the block timestamp method on the \emph{GoQuorum} (left) and \emph{Ganache} (right) local Ethereum blockchains for distinct blocktimes  $b=1,4,8$ (top to bottom).}
	\label{fig:localnetresults}
\end{figure*}

\subsection{Execution Accuracy}
\label{sec:executionaccuracy}

Accuracy is a standard metric used to evaluate the quality of a binary classifier in machine learning. It measures how well a binary classifier predicts the true class of an observation out of the two possible classes \emph{positive} and \emph{negative} correctly . Accuracy can be understood as the probability that a binary classifier correctly predicts the class of an observation to be positive when it is truly positive (true positive), and negative if it is truly negative (true negative). For a test set of $N$ observations, accuracy ${\cal A}$ of a binary classifier is defined as 
\begin{equation}\label{eq:accuracy}
	{\cal A} = (\text{true positives + true negatives})/N
\end{equation}
We now translate this accuracy metric for binary classifiers into the already outlined execution accuracy metric by establishing a link to the four distinct states of a time interval-constrained smart contract as shown in Table \ref{tab:confusiontable}.

We interpret the time-sensitive smart contract ${\cal C}$ as a binary classifier. From the perspective of the blockchain, we interpret observations as instances of injected time $\hat{t}_x$. Then, the state variable $p$ represents the true class of an observation and the state variable $\hat{p}$ represents the predicted class of an observation. More specifically, the state value $1$ is associated with the \emph{positive} class and the state value $0$ is associated with the \emph{negative} class. For a set $X$ of $n$ transactions, we define execution accuracy ${\cal A}_{\text{execution}}$ as 
\begin{equation}\label{eq:executionaccuracy}
	\begin{split}
	{\cal A}_{\text{execution}} &=\{ x \in X | t_x,\hat{t}_x \in I \lor  t_x,\hat{t}_x \notin I  \}/n\\
	&= \{ x \in X | {\cal C} \text{ is true positive}\}/n\\
	&\hspace{2cm}+ \{ x \in X | {\cal C} \text{ is true negative}\}/n	
	\end{split}
\end{equation}
in analogy to accuracy ${\cal A}$ in  Equation (\ref{eq:accuracy}). Observe that execution accuracy depends on the contraint interval $I$. Since $I$ is part of  the time-sensitive execution logic of ${\cal C}$, we see in particular that execution accuracy depends on ${\cal C}$'s  execution logic as desired. We now present empirical measurements of execution accuracy on four distinct Ethereum blockchains. 

\begin{figure*}[t]
	\centering
	\includegraphics[width=\linewidth]{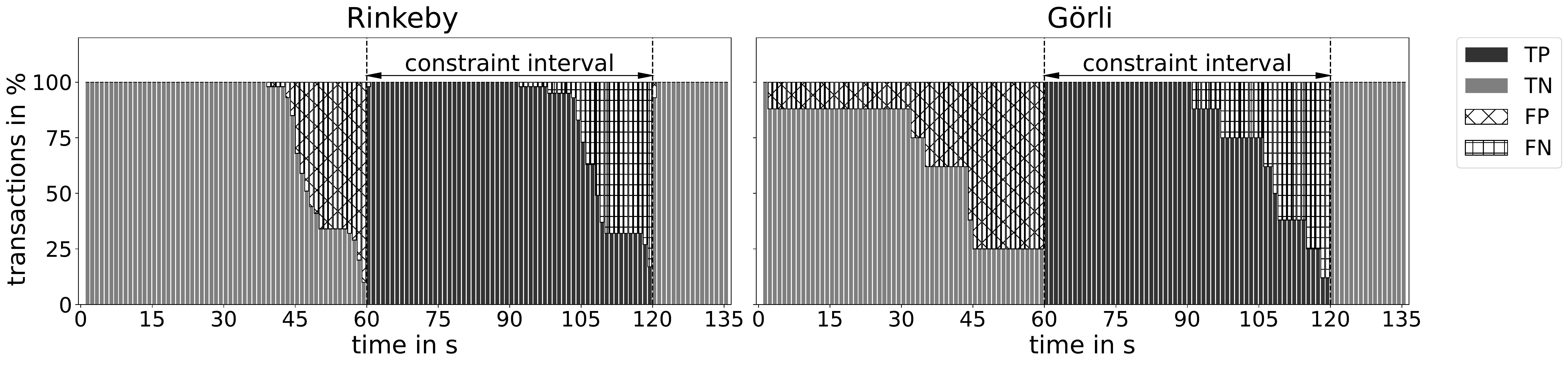}
	\caption{Relative frequencies of true positives (TP), true negatives (TN), false positives (FP), and false negatives (FN) for a time interval-constrained smart contract and the block timestamp method on the \emph{Rinkeby} (left) and \emph{G\"{o}rli} (right) Ethereum test networks.}
	\label{fig:testnetresults}
\end{figure*}

\section{Evaluation}
\label{sec:evaluation}

We measure execution accuracy of a time interval-constrained smart contract on the \emph{Ganache} and \emph{Quorum} local Ethereum blockchains and the \emph{G\"{o}rli} and \emph{Rinkeby} Ethereum test network. We will see that execution accuracy can behave differently on else equal configuration parameters. We make our source code available to the community.\footnote{\url{https://github.com/marcelTUB/Execution-Accuracy-Testbed}}

\subsection{Experimental Setup}
\label{sec:setup}

We implement a test bed following the description in Section \ref{sec:experimentalsetup} and a time interval-constrained smart contract ${\cal C}$ following the description in Section \ref{sec:smartcontract}. More specifically, we use constraint intervals of $I_{\text{local}}=[15,30]$, $I_{\text{test}}=[60,120]$ and experiment intervals \mbox{$J_{\text{local}}=[0,45]$}, \mbox{$J_{\text{test}}=[0,135]$} for the local and test networks respectively, where we measure time in seconds. We inject time via the block timestamp method as described in Section \ref{sec:notation}. We make available world time for smart contract execution via the parameter method. More specifically, we attach a string representing the current world when sending a transaction to the transaction.

 For each of the four blockchains to measure, we run 30 experiments. Over the course of an experiment,we send one transaction per second. After all transactions are mined, we read the now filled array holding the state changes $[p,\hat{p}]$.   Recall that  $p$ and $\hat{p}$ are the two binary state variables of ${\cal C}$ as shown in Table \ref{tab:confusiontable}. There, we see that ${\cal C}$ enters one of four distinct states after being called by a transaction: true positive ($p=1;\text{ }\hat{p}=1$), false positive ($p=0;\text{ }\hat{p}=1$), true negative ($p=0;\text{ }\hat{p}=0$), and false negative ($p=1;\text{ }\hat{p}=0$). If the array only holds true positives and true negatives, execution accuracy is perfect (see Equation (\ref{eq:executionaccuracy})). 
  
We will see in the following sections that the relative frequency with which a calling transaction $x$ changes the state of the smart contract ${\cal C}$ to any of the four states depends on the time $t_x$ of its sending. Note that the blockchains we measure use integer-valued block timestamps that represent full seconds. We therefore split results by their time of sending $t_x$  into batches of full seconds. Consider for instance the experiment interval $J=[0,3]$. We then split results into three batches of transactions sent during the subintervals $[0,1)$, $[1,2)$, and $[2,3]$ respectively. Note that sending many transactions within a short timeframe is prohibitive on most test blockains. We therefore average results per batch over multiple rounds of experiments per blockchain. 

\subsection{Execution Accuracy on Local Blockchain Networks}
\label{sec:localnetworkresults}

On local blockchain networks, measuring the impact of distinct parameter configurations on execution accuracy is readily feasible, as setting up a newly configured blockchain comes with little effort. We thus experimented with many distinct parameter configurations locally. We find that varying the \emph{blocktime} and  \emph{recommit interval length} has an immediate impact on execution accuracy. 

The blocktime defines the offset between block timestamps of consecutive blocks. Recommit intervals define periods of time when blockchain nodes aggregate transactions submitted to the blockchain. Aggregating transactions before they are presented to miners is usually beneficial from a cost perspective, since less candidate blocks are attempted  to be validated  by miners overall. Certainly, the cost aspect only holds true for non-local blockchains, while the impact of aggregation on execution accuracy pertains to local blockchains.

As expected, we find that increasing either the blocktime or the length of recommit intervals jeopardizes execution accuracy on average. Since the recommit interval length parameter is not available on the \emph{Quorum} blockchain, we omit reporting detailed results on changing the recommit interval length. Instead, we report results on varying the blocktime.

Figure \ref{fig:localnetresults} shows relative frequencies with which the smart contract ${\cal C}$ enters its four distinct states. Results are batched into batches of length 1 as described in Section \ref{sec:setup}. At large, both \emph{Ganache} and \emph{Quorum} yield similar results. We can see that the relative frequencies differ before, during, and after the constraint interval. We thus describe each segment separately.  

\textbf{Before Constraint Interval ($t_x \in [0,15)$):} To the left of constraint intervals, we only observe true negatives and false positives. Recall that execution accuracy equals the sum of the relative frequencies of true positives and true negatives (see Equation (\ref{eq:executionaccuracy})). Since the relative frequency of true positives is zero in this segment, execution accuracy is immediately equal to the relative frequency of true negatives. We thus see that execution accuracy is initially perfect and deteriorates approaching the left bound of the constraint interval. A comparison of \emph{Ganache} and \emph{Quorum} yields that \emph{Quorum} is more execution accurate than \emph{Ganache}. 

\textbf{During Constraint Interval ($t_x \in [15,30)$):} Within the constraint interval, we only observe true positives and false negatives. Inversely to the segment before the constraint interval, we have that execution accuracy is equal to the relative frequency of true positives instead of true negatives. Execution accuracy generally becomes lower near the interval bound. A comparison of \emph{Ganache} and \emph{Quorum} yields that \emph{Quorum} is more execution accurate than \emph{Ganache} at the lower interval bound yet less accuracte at the upper interval bound. 

\textbf{Past Constraint Interval ($t_x \in [30,45)$):} To the right of the constraint interval, we observe true negatives and false positives. Similarly to the segment before the constraint interval, we again have that execution accuracy equals the relative frequency of true negatives. Execution accuracy is lower near the upper interval bound and increases to perfect execution accuracy. A comparison of \emph{Ganache} and \emph{Quorum} yields that \emph{Quorum} is more execution accurate than \emph{Ganache}.
     
In summary, execution accuracy is generally perfect within and without the constraint interval. However, execution accuracy decreases toward interval bounds. This descrease is asymmetric,  that is execution accuracy is better past an interval bound than before. The extent of this discrepancy depends largely on the blocktime, yet also on the blockchain implementation at hand. We now measure the two test networks \emph{Rinkeby} and \emph{Goerli}, which we cannot configure at will.

\subsection{Execution Accuracy on Test Networks}
\label{sec:testnetworkresults}

Figure \ref{fig:testnetresults} shows relative frequencies for the \emph{Rinkeby} and \emph{G\"{o}rli} test networks. We see that the behavior of relative frequencies resembles that measured on local blockchains at large. Observe however that the \emph{G\"{o}rli} test network exhibits a lengthy period of false negatives in the interval $[0,30)$. An analysis yields that this is due to the \emph{G\"{o}rli} network sometimes having unexpectedly long response times. Transactions are sent long before the constraint interval, yet still found to satisfy the time interval constraint as they are presented and mined belatedly during the time of the constraint interval. We conclude that the \emph{Rinkeby} network is more execution accurate than the \emph{G\"{o}rli} network on average. 

\section{Conclusion}
\label{sec:conclusion}

The stipulation of time constraints via smart contracts on blockchains is inherently inaccurate. To date, this inaccuracy has only been characterized as the result of protocol and network delays and latencies irrespective of smart contract execution logic. We extend this characterization by proposing execution accuracy, a novel metric that quantifies this inaccuracy on the basis of smart contract execution logic instead. 

We specifically study interval time-constrained smart contracts that execute distinct logic within and without a pre-defined time interval. This class of smart contracts can for instance implement time-sensitive access control to a manufacturing device. Access is granted within a time interval, and else denied. Here, execution accuracy is the probability that access is accurately granted and denied. In contrast, inaccurate execution behavior encompasses cases in which access is falsely granted when requested without the interval, or falsely denied when requested within the interval. Our analysis confirms that execution accuracy decreases near interval bounds and in addition to prior work quantifies this decrease. 

The scope of the paper at hand is limited in three aspects. First, we only study absolute constraint intervals, that is constraint intervals that have fixed bounds. The study of execution accuracy for dynamic interval bounds is due. Second, we only study stateless execution logic, that is execution logic that is independent from the state of a smart contract. The study of stateful execution logic is due. Third, we only study the block timestamp time injection method. In particular, measuring execution accuracy of oracle-based time injection is due.

\newpage

\bibliographystyle{IEEEtran}
\bibliography{mainbib}

\end{document}